\documentclass[12pt,preprint]{aastex}
\slugcomment{To appear in the Astronomical Journal}

\lefthead{Skinner et al.}
\righthead{Wolf-Rayet Stars}

\begin{document}

\title{New X-ray Detections of WNL Stars}

\author{Stephen L. Skinner}
\affil{Center for Astrophysics and Space Astronomy (CASA), 
       Univ. of Colorado, 
       Boulder, CO 80309-0389; email: Stephen.Skinner@colorado.edu }

\author{Svetozar A. Zhekov}
\affil{Space and Solar-Terrestrial Research Institute, Moskovska str. 6,  
       Sofia-1000, Bulgaria}

\author{Manuel G\"{u}del}
\affil{Dept. of Astronomy, Univ. of Vienna, 
       T\"{u}rkenschanzstr. 17,  A-1180 Vienna, Austria}

\author{Werner Schmutz}
\affil{Physikalisch-Meteorologisches Observatorium Davos (PMOD),
       Dorfstrasse 33, 
       CH-7260 Davos Dorf, Switzerland}

\author{Kimberly R. Sokal}
\affil{Dept. of Astronomy, Univ. of Virginia, P.O. Box 400325,
       Charlottesville, VA 22904-4325}

%
\newcommand{\ltsimeq}{\raisebox{-0.6ex}{$\,\stackrel{\raisebox{-.2ex}%
{$\textstyle<$}}{\sim}\,$}}
%
\newcommand{\gtsimeq}{\raisebox{-0.6ex}{$\,\stackrel{\raisebox{-.2ex}%
{$\textstyle>$}}{\sim}\,$}}
\begin{abstract}
Previous studies have demonstrated that putatively single nitrogen-type Wolf-Rayet 
stars (WN stars) without known companions are X-ray sources. 
However, almost all WN star X-ray detections so far have been 
of earlier WN2 - WN6 spectral subtypes. Later WN7 - WN9 subtypes
(also known as WNL stars) have proved more difficult
to detect, an important exception being WR 79a (WN9ha).
We present here new X-ray detections of the WNL
stars WR 16 (WN8h) and WR 78 (WN7h). These new results, when combined 
with previous detections, demonstrate that X-ray emission is present in 
WN stars across the full range of spectral types, including later WNL stars.
The two  WN8 stars observed to date (WR 16 and WR 40) show unusually low 
X-ray luminosities (L$_{x}$) compared to other WN stars, and it is noteworthy 
that they also have the lowest terminal wind speeds (v$_{\infty}$).  
Existing X-ray detections of about a dozen WN stars reveal a trend of increasing
L$_{x}$ with wind luminosity L$_{wind}$ = $\frac{1}{2}$$\dot{\rm M}$v$_{\infty}^2$,
suggesting that wind kinetic energy  may play a key role in
establishing X-ray luminosity levels in WN stars.

\end{abstract}


%
\newpage

\section{Introduction}
The detection  of X-ray emission from Wolf-Rayet (WR) stars 
was one of the many notable discoveries of the {\em Einstein Observatory} 
(Seward et al. 1979; Pollock 1987). Subsequent studies have broadened our knowledge
of the X-ray properties of WR stars but the underlying physical 
mechanisms ultimately responsible for their X-ray emission are 
still not well understood. Most current  theoretical models 
attribute the X-ray emission of WR stars (and O-type stars) to 
thermal plasma produced in shocks that form   in their powerful
supersonic winds. WR stars have typical terminal wind speeds 
v$_{\infty}$ $\sim$ 1000 - 5000 km s$^{-1}$ 
and mass-loss rates $\dot{\rm M}$ $\sim$ 10$^{-5}$ M$_{\odot}$ yr$^{-1}$.
It has also been suggested that WR stars might   produce nonthermal 
X-rays via such processes as inverse-Compton scattering, synchrotron
emission, or nonthermal bremsstrahlung (Pollock 1987; Chen \& White 1991).
However, most WR stars detected so far show line-dominated X-ray
spectra that can be acceptably fitted by thermal models. But, nonthermal
emission has not yet been ruled out in a few cases (e.g. WR 142; Sokal
et al. 2010).

Several different shock-related processes could in principle lead to
thermal X-ray emission in WR stars. First, soft X-rays (kT  $<$ 1 keV)
may be produced in shocks embedded in the outflowing wind that form
as a result of the line-driven instability (Lucy \& White 1980; Lucy 1982; 
Feldmeier et al. 1997; Gayley \& Owocki 1995). This mechanism may be
responsible for the soft X-ray emission detected in some O-type stars
(Cassinelli et al. 2001; Kahn et al. 2001). But,  all WR stars detected
so far show a harder component (kT  $\gtsimeq$ 2 keV) in addition to a 
soft component that is usually present.  The hard component is not expected 
if the emission  arises solely from line-driven instability shocks.
Second, a sufficiently strong magnetic field could confine the wind
and channel it into two opposing streams which collide near the magnetic equator, 
forming a magnetically-confined  wind shock (MCWS). 
This mechanism is capable of producing hard X-rays (kT $\gtsimeq$ 2 keV)
and may be relevant to X-ray production in  young magnetic OB stars 
such as $\theta^{1}$ Ori C (Babel \& Montmerle 1997b; Gagn\'{e} et al. 2005)
and $\sigma$ Ori E (Skinner et al. 2008). However, very strong surface magnetic 
fields would be needed to confine the more powerful  winds of WR stars
(ud-Doula \& Owocki 2002;  Skinner et al. 2010, hereafter S10), and
observational evidence for strong  magnetic  fields on WR stars is currently
lacking. Thus, the question of whether the MCWS  mechanism operates in WR
stars remains open.  Finally, in the case of WR $+$ OB binary
systems, hard X-rays (kT $\gtsimeq$ 2 keV) can originate in the region
between the two stars where the interaction of their winds
results in a colliding wind (CW) shock (Prilutskii \& Usov 1976;
Luo et al. 1990; Stevens et al. 1992; Usov 1992). Hot X-ray
plasma that is thought to arise at least partially in a CW 
shock has now been detected in grating spectra of several 
X-ray bright WR $+$ OB binaries 
($\gamma^2$ Vel: Skinner et al. 2001; Schild et al. 2004;
WR 140: Pollock et al. 2005; WR 147: Zhekov \& Park 2010).  

X-ray spectra of WR stars are needed to distinguish between the above 
possibilities.   Although good-quality X-ray grating spectra  of a few 
X-ray bright WR$+$OB binaries have been obtained (as noted above), 
the study of putatively single WR stars as a class has been  hampered 
by a general lack of X-ray spectral data. The observational situation
has improved considerably during the past few years as a result of  the 
completion  of an exploratory  X-ray survey of single WR stars with {\em XMM-Newton}
and {\em Chandra} (S10). Moderate-resolution
CCD X-ray spectra are now available for several WN stars but attempts
to detect carbon-rich WC stars have so far only yielded upper limits
on their X-ray luminosities with typical constraints of 
log L$_{x}$ $<$ 31.0 ergs s$^{-1}$ (Skinner et al. 2006; see also Fig. 9 of S10).

A surprising result of the initial survey was that all detected WN
stars showed evidence of hot plasma (kT$_{hot}$ $\gtsimeq$ 2 keV) in their
X-ray spectra. This result is at odds with X-ray production via
the line-driven instability mechanism, for which only soft emission
is expected. It is thus apparent that other
mechanisms are needed to fully explain the origin of X-ray emission
in WN stars and, within the current theoretical framework,
either the MCWS or CW shock  processes could be responsible. 
However, there is no compelling justification to invoke the CW 
interpretation because the WN stars in the initial survey sample 
were selected on the basis of lack of evidence for binarity.
The presumption that these stars are indeed single  is of course 
subject to the usual caveat that close companions can escape
detection.

In the initial survey,  {\em XMM-Newton} and {\em Chandra}   
data for the following detected WN stars were analyzed:
WR 2 (WN2), WR 18 (WN4), WR 20b (WN6h),  WR 24 (WN6ha), 
WR 134 (WN6), WR 136 (WN6h), and WR 79a (WN9ha).
In addition, archival {\em ROSAT} data were analyzed
for WR 16 (WN8h) and WR 78 (WN7h). Neither of these two WNL stars 
\footnote{Nitrogen-rich WR stars with spectral types in the 
range WN7 - WN9 are referred to as late WN stars, or WNL stars.
Those with spectral types of WN2 - WN5 are called early WN
stars or WNE stars.  WN6 stars are intermediate and may
be classified as either WNE or WNL (Crowther 2007). A ``h'' suffix
in the spectral type denotes the presence of hydrogen emission
lines in the spectrum.}
showed significant X-ray emission (see also Sanders et al. 1985), 
although a marginal (2$\sigma$) excess was noted 
in hard-band {\em ROSAT} 
PSPC\footnote{The {\em ROSAT} Position Sensitive Proportional
Counter (PSPC) provided energy coverage in the $\approx$0.1 - 2.35  keV energy
range but lacked sensitivity at higher energies E $\gtsimeq$ 2.35 keV.
PSPC hard-band images cover the range E $\approx$ 0.5 - 2 keV.
In contrast, {\em XMM-Newton}  provides coverage to higher energies and 
we  use the energy range  E = 2 - 8 keV to refer to the {\em XMM-Newton} hard-band.}
images of WR 78. We present here more sensitive {\em XMM-Newton} observations of 
WR 16 and WR 78 that were obtained after the results of our  
initial WN survey were published (S10). These observations are 
important  because they provide additional information on
the X-ray properties of WNL subtypes, which comprise
$\approx$24\% of all WR stars listed in the {\em VIIth Catalog
of Wolf-Rayet Stars} (van der Hucht 2001).  WNL stars are believed
to be the least chemically-evolved WR stars and some O-type
stars probably enter the WR phase as WNL stars (Crowther \& Bohannan 1997).
WNL stars  typically have higher mass-loss rates but lower terminal wind speeds 
than WNE stars (Table 2 of  Crowther 2007).
We report here a faint X-ray detection of WR 16 (WN8h)  and bright emission 
including a hot plasma component from  WR 78 (WN7h).

\section{Observations}

The {\em XMM-Newton} observations are listed  in Table 1.
The general properties of the WNL stars discussed here are
summarized in Table 2 and their X-ray properties are given in
Table 3.  The targets were selected from 
the {\em VIIth WR Catalog}  (van der Hucht 2001)
based on proximity, low visual extinction, and lack of
evidence for binarity. In addition to the new detections
of WR 16 and WR 78, we also include data from our previous
{\em XMM-Newton} observation (S10) of the WNL star WR 79a  
for comparison.

Data were acquired with the European Photon
Imaging Camera (EPIC) in full-window mode. EPIC provides 
charge-coupled device (CCD) imaging spectroscopy from the 
pn camera (Str\"{u}der et al. 2001) and two nearly
identical MOS cameras (MOS1 and MOS2; Turner et al. 2001).  
The EPIC cameras provide  energy coverage in the range
E $\approx$ 0.2 - 15 keV with energy 
resolution E/$\Delta$E $\approx$ 20 - 50.
The MOS cameras provide the best on-axis angular 
resolution with FWHM $\approx$ 4.3$''$ at 1.5 keV.

Data were reduced using the {\em XMM-Newton}
Science Analysis System (SAS vers. 11.0) using
standard procedures including the filtering of
raw event data to select good event patterns
and removal of data within time intervals of
high background radiation. 
Spectra and light curves were extracted 
from  a  circular region of radius r = 15$''$
($\approx$68\% encircled energy) centered on the source,
thus capturing most of the source 
counts while reducing background counts. Background
analysis was conducted on circular source-free regions near
the source. The SAS tasks {\em rmfgen} and {\em arfgen} were 
used to generate source-specific RMFs and ARFs for spectral 
analysis. The data were analyzed using the HEASOFT {\em Xanadu}
software package.

\section{Results }

\subsection{WR 16 (= V396 Car = HD 86161)}
This WN8h star has the lowest terminal wind speed (v$_{\infty}$ $\approx$
630 - 740 km s$^{-1}$) of 
any WN star in our exploratory X-ray survey. The star is surrounded by a cocoon
of molecular gas which was probably ejected during a previous
evolutionary phase (Marston et al. 1999).  Optical variability
is present but its cause is not fully understood (Antokhin et al. 1995).
An early study attributed the variability to an unseen low-mass
companion (Moffat \& Niemela 1982) but this interpretation was
later called into question   (Manfroid et al. 1987).
The stellar distance is somewhat uncertain with published
values ranging from 2.37 - 4.36 kpc (Crowther et al. 1995;
Nugis \& Lamers 2000; van der Hucht 2001, Hamann et al. 2006) and we 
adopt here an  intermediate value d $\approx$ 3.6 kpc.

Visual inspection of energy-filtered EPIC images showed no
obvious signs of an X-ray source at the stellar optical position.
However, more in-depth image analysis revealed a weak emission
excess above the local background level in both the pn and MOS
images. We extracted  counts (0.3 - 8 keV) within a
circular region of radius 15$''$ (68\% encircled energy) 
centered on the star in each EPIC detector, and background counts 
in different source-free regions near the star. After background 
subtraction we obtained 31$\pm$6 net pn counts in 30.0 ks of usable
exposure and 24$\pm$5 net MOS1$+$2  (sum of MOS1 and MOS2) in 66.0 ks
of usable exposure (33.0 ks per MOS). The faint emission is clearly
visible in the contoured MOS1$+$2 image in Figure 1. Although the 
statistical significance of the excess is low ($\approx$2$\sigma$),
its presence in all three EPIC detectors at an offset of $<$1$''$
from the {\em HST} GSC optical position (Table 3) strengthens
the conclusion that weak X-ray emission from WR 16 is detected.

The emission is too faint for spectral and timing analysis.
However, we used the PIMMS \footnote{For information 
on PIMMS (Portable Interactive Multi-Mission Simulator) 
see http://cxc.harvard.edu/ciao/ahelp/pimms.html .}
simulator to estimate
the X-ray luminosity of WR 16 based on the count rates.
The pn rate of 1.03 c ks$^{-1}$ gives an unabsorbed luminosity
log L$_{\rm X}$(0.3 - 8 keV)  = 31.14 [30.77 - 31.30] ergs s$^{-1}$ 
for an assumed distance d = 3.6 kpc, where the range in 
square brackets reflects the range of published distances d = 2.37 - 4.36 kpc.
A PIMMS estimate based on the MOS count rate 
gives a nearly identical unabsorbed value 
log L$_{\rm X}$(0.3 - 8 keV)  = 31.17 [30.80 - 31.33] ergs s$^{-1}$.
For the underlying spectrum, we assumed an absorbed
two-temperature optically thin thermal plasma
($apec$) with kT$_{1}$ = 0.6 keV and 
kT$_{2}$ = 3.5  keV, typical of detected WN stars (S10).
We adopted an absorption
column density N$_{\rm H}$ = 3.7 $\times$ 10$^{21}$ 
cm$^{-2}$ based on A$_{\rm V}$ = 1.67 (Table 1) and
the Gorenstein (1975) conversion from A$_{\rm V}$ 
to N$_{\rm H}$. The slightly different conversion of
Vuong et al. (2003) decreases the  upper limit on L$_{\rm X}$ 
by 0.05 dex. 

The above  L$_{\rm X}$ is the lowest of any WN star detected so far,
but the derived value is dependent on the assumed spectral model
and absorption. Most WN stars show X-ray absorption   in
excess of that expected from A$_{\rm V}$ (S10; see also the 
results for WR 78 in Sec. 3.2 below). If that is the
case for WR 16, then its unabsorbed L$_{\rm X}$ could be larger
than the value inferred above. In this regard, we note that the
pn counts for  WR 16 are distributed roughly equally between the soft (0.3 - 2 keV)
and hard (2 - 8 keV) bands. But, XSPEC simulations show that if the 
absorption were as low as expected based on A$_{\rm V}$, only about
$\approx$15\% - 20\% of the counts would emerge in the hard band.  
This suggests that excess absorption above that expected from
A$_{\rm V}$ is present in WR 16, but a higher signal-to-noise 
spectrum would be needed to confirm this. For those WN stars where
sufficient counts are available for spectral analysis, the best-fit
N$_{\rm H}$ values are typically $\approx$2.0 - 2.5 times larger than expected 
from estimates based on A$_{\rm V}$ (Table 4 of S10). If we assume
that is the case for WR 16 then its unabsorbed L$_{\rm X}$ determined from 
PIMMS is $\approx$0.10 - 0.15 dex larger than the values quoted above.

\subsection{WR 78 (= V919 Sco = HD 151932)}

This WN7h star has been included in numerous optical studies
and there is fairly  good agreement in published stellar 
parameters when different studies are compared (Table 2).
Optical variability has been reported (Vreux et al. 1987; Bratschi \& Blecha 1996)
but no periodicity that would  hint of a companion has been found.

{\em XMM-Newton} provides the first unambiguous detection of 
WR 78 (Fig. 2). The source is quite hard as is apparent
from its median photon energy E$_{50}$ = 2.47 keV (Table 3).
Clear evidence for high-temperature plasma is provided by
the detection of the  Fe K$\alpha$ emission line complex
at 6.67 keV (Fig. 3). This line forms at a characteristic
temperature  T $\sim$ 40 MK. Using pn events in the 0.3 - 8 keV
range, the Kolmogorov-Smirnov (KS) test gives a probability of 
constant count rate  P$_{cons}$ = 0.12 and a $\chi^2$ test 
applied to the background-subtracted pn light curve binned at 
1000 s intervals gives P$_{cons}$ = 0.97. The KS test operates
on all detected events within the source extraction region, 
and thus includes both background and source events. The lower
value of P$_{cons}$ based on the KS test may be due to 
background effects. But, neither the KS test nor $\chi^2$
analysis suggests statistically significant variability for
WR 78 during the 26.26 ks (7.3 hours) of usable exposure.
Likewise, the other WN stars in the survey have failed to 
show any clear X-ray variability in observations spanning
$\approx$3 - 10 hours (S10).

Acceptable fits of the CCD X-ray spectra of WN stars detected in 
the initial survey  were obtained using a two-temperature (2T)
optically thin plasma model consisting of 
cool (kT$_{1}$) and hot (kT$_{2}$) plasma components plus a single
absorption component (N$_{\rm H}$). This model is expressed 
symbolically as N$_{\rm H}$$*$(kT$_{1}$ $+$ kT$_{2}$), where  
parentheses denote that both plasma components are viewed
through the same absorption column. Interestingly, this model
gives an unacceptable fit of the WR 78 X-ray spectrum 
as does a simpler 1T model of the form N$_{\rm H}$$*$kT$_{1}$.   
These single-absorption models fit the spectrum above 2 keV
(including the Fe K$\alpha$ line) reasonably well, but
generally underestimate the observed flux at energies
below 1 keV. A similarly poor fit below 1 keV occurs for
other single-absorption models such as a plane-parallel
shock model ($vpshock$ in XSPEC). In order to obtain an 
acceptable fit for WR 78, different absorption components 
were required for the cool and hot components, i.e. a model of the form
N$_{\rm H,1}$$*$kT$_{1}$ $+$ N$_{\rm H,2}$$*$kT$_{2}$.

The acceptable two-absorber  model is summarized in Table 4. 
The inferred plasma temperatures kT$_{1}$ = 0.64 [0.48 - 0.88; 90\% conf.] keV 
and kT$_{2}$ = 2.27 [1.63 - 2.90] keV are not substantially different
from the WN9h star WR 79a (Sec. 3.3) when 90\% confidence ranges are
taken into account, but the value of kT$_{2}$ is at the low end of
the range for all WN stars observed so far (S10).
Most of the X-ray flux and emission measure (as gauged by the value of
$norm$ in Table 4) of WR 78 is associated with the hotter plasma component,
which is evidently viewed through much higher absorption than
the cool component. The separate contributions of the cool and
hot components are shown in the bottom panel of Figure 3.
 The best-fit absorption values of 
N$_{\rm H,1}$ = 1.38 [1.20 - 1.60] $\times$ 10$^{22}$ cm$^{-2}$ and 
N$_{\rm H,2}$ = 6.55 [4.10 - 10.7] $\times$ 10$^{22}$ cm$^{-2}$
are substantially larger than expected on
the basis of visual extinction esimates, since 
A$_{\rm V}$ = 1.55 mag (Table 2) corresponds to N$_{\rm H}$ =
3.4 $\times$ 10$^{21}$ cm$^{-2}$ using the conversion 
of Gorenstein (1975). The two-absorber model for
WR 78 gives an unabsorbed X-ray luminosity log L$_{x}$(0.3 - 8 keV) =
32.84 ergs s$^{-1}$ (d = 1.99 kpc; Table 4), which is at the
high end of the range for detected WN stars (Fig. 4; S10). However,
this value is quite uncertain because of the large dereddening
correction. All spectral models that we considered, including
simplistic 1T models, give values
log L$_{x}$(0.3 - 8 keV) $\geq$ 32.1 ergs s$^{-1}$
which we consider to be a reliable lower limit.

The most likely cause of the anomalous  X-ray absorption is the 
WR wind. In that case, the higher absorption associated with 
the hot plasma component suggests that it originates deeper in the
wind (closer to the star) where the wind density is higher.
A similar conclusion was reached for the WN6h star WR 136
(S10) and for O-type stars studied by
Waldron \& Cassinelli (2007) and Naz\'{e} (2009).  In the ideal case of a 
smooth unclumped wind,  the radius of optical
depth unity (as measured from the star) decreases toward
higher X-ray energies so harder  photons are able to
escape from smaller radii near the star. Softer X-ray photons
produced far out in the wind suffer less wind absorption
and can escape whereas any soft photons produced in deeper
denser layers of the wind are absorbed and remain undetected.
In the more complex case of an  inhomogenous (clumped) wind, 
the above picture is not altered substantially if the clumps 
are optically thin to X-rays but could well be different if
dense optically thick clumps are present.

\subsection{WR 79a (= HD 152408) }
This WN9ha star is the latest WN subtype 
detected in X-rays so  far. It is discussed in more
detail in S10 but for comparison  we summarize its  
X-ray properties in Table 3 and give its best-fit 
spectral parameters in Figure 3. The EPIC pn spectrum
of WR 79a is softer than that of WR 78 as evidenced
by its lower median photon energy (Table 3) and it
is apparent from Figure 3 that the X-ray absorption
of WR 79a is less than that of WR 78. The pn spectra
of both stars show the high-temperature S XV (2.46 keV)
emission line but the very high temperature 
Fe K$\alpha$ complex (6.67 keV)  detected
in WR 78 is not seen in WR 79a.

\section{Discussion}

\subsection{What Stellar Parameters Control WN Star X-ray Luminosity?}
The fundamental stellar parameters that govern intrinsic X-ray 
luminosity levels in WR stars have not yet been identified.
Good quality X-ray data or useful upper limits exist  for only
about a dozen putatively single WN stars, so robust statistics 
and stringent tests for correlations of L$_{x}$ against stellar 
parameters are still limited by small sample size. This is in 
contrast to massive O-type stars, from which WR stars evolve.
X-ray data now exist for hundreds of O-type stars and 
the empirical  relation L$_{x}$ $\sim$ 10$^{-7}$L$_{bol}$
was already becoming apparent from a large sample of O and early B stars
detected in the {\em ROSAT} All-Sky Survey (Bergh\"{o}fer et al. 1997).

Based on existing {\em XMM-Newton} and {\em Chandra} X-ray data, there 
is no convincing  evidence for a similar  L$_{x}$ $\propto$ L$_{bol}$ relation 
in putatively single WN stars. Likewise, no evidence for such a relation
was found using {\em ROSAT} data (Wessolowski 1996). As we have
already shown (S10), WN stars with similar values of
L$_{bol}$ can have L$_{x}$ values differing by an order of magnitude
or more. Most detected WN stars, including WR 78 (Table 4), 
have L$_{x}$ $\gtsimeq$ 10$^{-7}$L$_{bol}$ with typical values 
being L$_{x}$ $\sim$ 10$^{-6.5}$L$_{bol}$, as shown in Figure 4. 
However there are a few WNL stars whose   L$_{x}$/L$_{bol}$ ratios 
are at least an order of magnitude lower. Most notable among these
is WR 16 (WN8h), for which we obtain 
log L$_{x}$/L$_{bol}$ = $-$8.2 $\pm$ 0.2 when taking the range of
published distances and stellar luminosities into account (Table 2).
A similarly low L$_{x}$/L$_{bol}$ ratio is also assured for WR 40 (WN8h)
by virtue of its {\em XMM} non-detection with an  X-ray upper limit
log  L$_{x}$(0.5 - 10 keV)  $\leq$ 31.6 ergs s$^{-1}$ at an
assumed distance d = 3 kpc (Gosset et al. 2005).

Wind shock models of X-ray emission from massive stars
predict that  L$_{x}$ will depend on mass-loss parameters.
In the line-driven instability shock model,
L$_{x}$ $\propto$ $\dot{\rm M}^{\alpha}$ where
$\alpha$ = 1 for the radiative shock case,
$\alpha$ = 2 for the adiabatic case, and
$\alpha$ = 0.6 if the observed L$_{x}$ $\propto$
L$_{bol}$ scaling relation in O-type stars is to be reproduced
(Owocki et al. 2011). From an observational standpoint, there is no
clear dependence of L$_{x}$ on $\dot{\rm M}$ for WN
stars. In particular WR 2 (WN2) has one of the lowest
mass-loss rates tabulated for WN stars (log $\dot{\rm M}$ = $-$5.4 - $-$5.3
M$_{\odot}$ yr$^{-1}$; Nugis \& Lamers 2000; Hamann et al. 2006) 
but its L$_{x}$ is comparable to that of other WN stars
having $\dot{\rm M}$ values an order of magnitude larger.
In addition, the two WN8 stars WR 16 and WR 40
are underluminous in X-rays by about an order of magnitude
compared to other WN stars with similar mass-loss rates. 
Thus, mass-loss rate alone
is not a reliable predictor of L$_{x}$ in WN stars.


There is one clue in the existing data that suggests
L$_{x}$ in WN stars may depend  on terminal wind speed v$_{\infty}$
or some function thereof. Specifically, we note that the
two observed WN8 stars WR 16  (v$_{\infty}$ $\approx$
630 - 740  km s$^{-1}$)
and WR 40 (v$_{\infty}$ $\approx$ 650 - 910 km s$^{-1}$)
have the lowest terminal wind speeds of all WN stars
(Hamann et al. 2006;  see also Gosset al. 2005 for WR 40).
These two stars also have the lowest L$_{x}$ values, even
though only an upper limit is currently  available for WR 40.
The WN8h star WR 156  also has a similar low wind speed 
(v$_{\infty}$ $\approx$ 660 km s$^{-1}$; Hamann et al. 2006)
but it  has not yet been observed in X-rays. In  the
absence of a binary companion we expect that any
X-ray detection of WR 156 would  be quite faint.
 
The apparent falloff in L$_{x}$ with decreasing v$_{\infty}$
suggests that  L$_{x}$ may be related to  
wind momentum (= $\dot{\rm M}$v$_{\infty}$) or wind luminosity
L$_{wind}$ (= $\frac{1}{2}$$\dot{\rm M}$v$_{\infty}^2$).
Using data in the initial WN X-ray survey (S10), we already
noted a trend of increasing L$_{x}$ with L$_{wind}$. The 
new detections of WR 16 and WR 78 presented here conform to 
and strengthen this trend (Fig. 5). This trend was not seen
in {\em ROSAT} WN-star data analyzed by Wesselowski (1996),
perhaps because {\em ROSAT} had little sensitivity above
$\sim$2 keV where a significant fraction of the X-ray flux
in WN stars emerges.
Using the data in Figure 5, the generalized Kendall's tau test gives
a correlation probability P$_{corr}$ = 0.93. A linear regression
fit gives log L$_{x}$ = 0.89($\pm$0.47)log L$_{wind}$ $+$ 26.41($\pm$0.55)
where L$_{x}$ is in units of ergs s$^{-1}$ and 
L$_{wind}$ in units of 10$^{-5}$M$_{\odot}$ yr$^{-1}$ $\times$ (km s$^{-1}$)$^2$.
Thus, the regression fit gives a nearly linear dependence L$_{x}$ $\propto$ 
L$_{wind}^{0.89\pm0.47}$ but the slope is quite  uncertain due to scatter 
in the data. Factors  that  could contribute to this scatter include uncertainties in 
distances, X-ray luminosity dereddening corrections, WR star mass-loss 
rates (which are sensitive to clumping effects), and terminal wind speeds.

A notable outlier is the WN6h star WR 136, which appears to be
underluminous in X-rays compared to other WN stars with comparable  L$_{wind}$ 
values (Fig. 5) and to other WN stars with comparable  stellar luminosities
L$_{star}$ (Fig. 4). Its L$_{star}$ value is also low compared to other WNh stars
(Fig. 6 of Hamann et al. 2006). This raises a question as to whether
current distance estimates may be too low. The estimates  range from
d = 1.26 kpc (Hamann et al. 2006) to d = 1.82 kpc (Nugis \& Lamers 2000),
whereas our derived X-ray luminosity log L$_{x}$ = 31.51 ergs s$^{-1}$ 
is based on the {\em Hipparcos} distance d = 1.64 kpc (S10). We note
that there is also a large uncertainty in the  L$_{x}$ value of WR 136
because of poorly-constrained X-ray absorption (N$_{\rm H}$) in spectral fits.
Some statistically-acceptable spectral models require higher absorption 
and yield dereddened values as high as log  L$_{x}$ $\approx$ 32.2 ergs s$^{-1}$
(d = 1.64 kpc). Such higher values, if correct, would shift the position
of WR 136 upward in Figures 4 and 5 (as shown by the  L$_{x}$ error bar),   
in better agreement  with other WN stars.

If the above L$_{x}$ $\propto$ L$_{wind}$  relation reflects a 
true physical dependence, then the obvious  conclusion would be that 
a relatively constant fraction of the kinetic energy in  WN star winds
is being converted into X-ray energy. Interestingly, the MCWS model
of Babel \& Montmerle (1997a) predicts such a dependence. Specifically,
they obtain 
L$_{x}$ $\approx$ $\frac{1}{2}$$\dot{\rm M}$v$_{shock}^2$, and
they infer v$_{shock}$ $\sim$ v$_{\infty}$ in order to 
reproduce the X-ray luminosity of the A0p star IQ Aur from 
their model. Thus, there is some theoretical support for a
L$_{x}$ $\propto$ L$_{wind}$ relation in early-type stars
with magnetic fields. Even so, some observational scatter in 
this relation would be expected due to differences in B-field 
strengths between stars (eq. [10] of Babel \& Montmerle 1997a) 
and bearing in mind that collision of the two hemispherical wind components 
could in some cases occur at subterminal speeds. Although
the existence of strong surface magnetic  fields on single WR stars 
is not yet established,  some WR $+$ OB binaries do exhibit nonthermal 
radio emission which may be associated with  weak magnetic
fields above the surface (Abbott et al. 1986; Skinner et al. 1999; 
Dougherty \& Williams 2000). Since the MCWS mechanism operates in single
stars  there is no need to postulate unseen companions in this
interpretation. 

If close unseen companions are present in our sample
of putatively single WN stars then hot plasma could  be created via
the colliding wind mechanism. If the WR wind momentum dominates that
of the companion, then the CW shock would form at or near the companion
surface. This could also give rise to a L$_{x}$ $\propto$  L$_{wind}$ 
dependence, but considerable scatter would be expected since
in this case the predicted  L$_{x}$ also depends on  binary separation
and the radius of the companion star (Usov 1992; Luo et al. 1990).
Since we have no specific information on the existence of such companions,
their stellar or wind properties, or their orbital parameters, a meaningful 
comparison of the observed L$_{x}$ values with CW model predictions 
cannot yet be made.


\subsection{High  X-ray Plasma Temperatures in WN Stars}

All WN stars for which good-quality CCD X-ray spectra exist
show a high-temperature plasma 
component (kT$_{hot}$ $\gtsimeq$ 2 keV) and almost always a 
cooler component (kT$_{cool}$ $<$ 1 keV). The cool
component could be due to radiative instability
wind shocks, but the origin of the hotter component is
not yet known. Analysis of {\em XMM} EPIC and {\em Chandra}
ACIS spectra of WN stars reveals a broad range of hot-component
temperatures from kT$_{hot}$ $\approx$ 2 keV (WR 1, WR 78) up to 
kT$_{hot}$ $\approx$ 4 - 8 keV  (WR 20b, WR 110, WR 134; S10). 

Such high-temperature plasma could  be produced by supersonic
WR winds shocking onto denser downstream material, or the 
wind from the opposite hemisphere  (as in the MCWS model), or the wind 
or surface of as yet undetected companions (CW shocks).   
These models all predict that the {\em maximum}  shock
temperature should scale proportionally with the pre-shock
wind-speed as kT$_{shock}^{(max)}$ $\approx$ 1.96$\mu$v$_{1k}^2$ keV,
where $\mu$ is the mean atomic weight per particle and
v$_{1k}$ is the wind speed in units of 1000 km s$^{-1}$
(Babel \& Montmerle 1997a; Luo et al. 1990;, Stevens et al. 1992).
For solar abundances $\mu$ $\approx$ 0.6 but for metal-enriched
WN star winds $\mu$ $\approx$ 1.06 (van der Hucht et al. 1986).
For the WN star case of interest here one thus obtains 
kT$_{shock}^{(max)}$ $\approx$ 2v$_{\infty,1k}$$^2$ keV, assuming
that the wind has reached terminal speed at the shock
interface.

Terminal wind speeds of X-ray detected WN stars lie in the range
$v_{\infty}$ $\approx$ 700 - 2500 km s$^{-1}$ (Table 2; S10).
Thus, the above shock models predict 
maximum shock temperatures kT$_{shock}^{(max)}$ $\approx$ 1 - 12 keV.
The values of kT$_{hot}$ determined from X-ray spectra lie
comfortably within this range. However, there is no clear trend for 
a proportionality relation  kT$_{hot}$ $\propto$ v$_{\infty}^2$,
as one might naively expect from shock-model predictions.
The WN star with the lowest v$_{\infty}$ for which a good-quality
CCD spectrum is available is the WN9ha star WR 79a 
(v$_{\infty}$ $\approx$ 955 km s$^{-1}$, Table 2) 
and its best-fit temperature is kT$_{hot}$ =  2.66 [1.66 - 5.60; 90\% conf.]  keV
(S10). By comparison, the WN4 star WR 1 has a higher terminal speed 
(v$_{\infty}$ $\approx$ 1600 - 2135  km s$^{-1}$; Nugis \& Lamers 2000;
Ignace et al. 2001; Hamann et al. 2006) but  a lower best-fit
temperature kT$_{hot}$ =  1.72 [1.36 - 2.21; 90\% conf.]  keV
based on our analysis of {\em XMM} archive data (obsid 0552220101;
see also Ignace et al. 2003a for analysis of an earlier {\em XMM} 
observation with a shorter exposure).

Given the above results, it is not yet clear that terminal wind
speed is the primary parameter governing hot-plasma component
temperatures in WN stars. Other factors such as magnetic fields
(if present) could play a role in plasma heating. Even if the
hot-component plasma temperature is governed by v$_{\infty}$,
establishing such a dependence from the existing data will 
not be straightforward. As noted above, shock models predict that the
maximum shock temperature kT$_{shock}^{(max)}$ $\propto$ v$_{\infty}^2$,
but shocks will in general produce plasma over a range of temperatures.
X-ray spectral fits of CCD spectra using simplistic 2T models
only provide  an estimate of the {\em average} hot-component temperature
kT$_{hot}$, which will in general  be less than kT$_{hot}^{(max)}$.
Furthermore, the inferred values of kT$_{hot}$ from fits of moderate
resolution CCD spectra can be uncertain by a factor of $\sim$2 (S10).
By comparison, values of v$_{\infty}$ are typically uncertain
by $\approx$10\% - 20\%. More accurate temperature discrimination 
is needed to search for  any correlation of maximum plasma temperature with
wind speed. High-resolution X-ray grating spectra can
provide more detailed information on the distribution of X-ray
plasma emission measure as a function of temperature. But improvements in X-ray
telescope sensitivity will be needed to acquire grating spectra
for a sample of WN stars that is sufficiently large to 
undertake correlation studies.

\section{Summary}

The main results of this study are the following:

\begin{enumerate}

\item We report new X-ray detections of the WNL stars WR 16 (WN8h)
      and WR 78 (WN7h). These new detections, when combined with 
      previous detections show that X-ray emission is present
      in all WN subtypes from WN2 - WN9. This contrasts sharply
      with single carbon-rich WC stars which so far remain undetected
      in X-rays.  

\item The {\em XMM} detection of WR 16 is of low significance and
      the emission is too faint for spectral analysis. However,
      the X-ray spectrum of WR 78 reveals a high-temperature plasma
      component (kT$_{hot}$ $\approx$ 2 keV), similar to that observed 
      in other WN stars. Such hotter plasma is not predicted for X-ray
      emission arising solely in line-driven instability shocks.
      The origin of the hot plasma in WN stars is not yet known.
      Shock models predict that the maximum  plasma temperature
      will scale with terminal wind speed according to
      kT$_{shock}^{(max)}$ $\propto$ v$_{\infty}^2$. Such a dependence
      is not yet seen in the existing data but could be masked because
      temperatures determined from CCD X-ray spectra are average
      (not maximum) values and are in many cases  uncertain by 
      a factor of $\sim$2.

\item The X-ray luminosities of the   WN8 stars WR 16 (a faint detection)
      and WR 40 (a non-detection)  are  an order of magnitude
      less than typical values observed for other detected WN stars.
      These WN8 stars also have the lowest terminal wind speeds of any WN 
      stars observed so far, suggesting a possible link between terminal 
      wind speed (or a parameter that depends on wind speed)
      and X-ray  luminosity in single WN stars.

\item The two new X-ray detections of WR 16 and WR 78 conform to and strengthen a previously
      noted trend of increasing L$_{\rm x}$ with  wind luminosity 
      L$_{wind}$ = $\frac{1}{2}$$\dot{\rm M}$v$_{\infty}^2$ in WN stars.
      A regression fit based on existing data for 11 WN stars
      gives L$_{x}$ $\propto$ L$_{wind}^{0.89\pm0.47}$ and a correlation
      test yields a correlation probability P$_{corr}$ = 0.93.
      This trend needs to be confirmed in a larger sample of WN stars but
      tentatively suggests that wind  kinetic energy may be an important
      factor  in governing the X-ray luminosity  levels of WN stars.

\end{enumerate}

\acknowledgments

This work was supported by  NASA/GSFC award NNX09AR25G. 
This work was  based on observations obtained with 
{\em XMM-Newton}, an ESA science mission with instruments and
contributions directly funded by ESA member states
and the USA (NASA). SZ acknowledges financial support
from Bulgarian National Science Fund grant DO-02-85. 

\clearpage

%
%
%
\begin{deluxetable}{lll}
\tabletypesize{\scriptsize}
\tablewidth{0pc}
\tablecaption{XMM-Newton Observations (WNL Stars) }
\tablehead{
\colhead{Parameter} &
\colhead{WR 16} &
\colhead{WR 78} \\
}
\startdata
Date              & 2009 Dec. 28-29  & 2010 Mar. 2-3    \nl
ObsId             & 0602020301       & 0602020201       \nl
Start Time (UTC)  & 19:57:31         & 14:08:02         \nl
Stop  Time (UTC)  & 06:35:01         & 00:45:08         \nl
Optical Filter    & medium           & thick            \nl
pn exposure (ks)\tablenotemark{a}  & 30.0             & 26.3        \nl
MOS exposure (ks)\tablenotemark{a} & 33.0             & 26.5        \nl
\enddata
\tablenotetext{a}{Usable exposure after removing high-background time intervals. MOS exposure
                  is per MOS.  The total exposure times before removing high-background intervals
                  were 37.4 ks (WR 16), 37.5 ks (WR 78).}
\end{deluxetable}
\clearpage
\begin{deluxetable}{llccccc}
\tabletypesize{\scriptsize}
\tablewidth{0pc}
\tablecaption{WNL Star Properties\tablenotemark{a} }
\tablehead{
\colhead{Name} &
\colhead{Sp. Type } &
\colhead{d} &
\colhead{A$_{\rm V}$ } &
\colhead{v$_{\infty}$} &
\colhead{log $\dot{\rm M}$} &
\colhead{log L$_{star}$} \\
\colhead{         } &
\colhead{         } &
\colhead{(kpc)} &
\colhead{(mag)} &
\colhead{(km/s)} &
\colhead{(M$_{\odot}$/yr)} &
\colhead{(L$_{\odot}$)}  \\
}
\startdata
WR 16  & WN8h    & 3.60 (2.37 - 4.36)  & 1.67 & 670 (630 - 740)    & $-$4.35 ($-$4.55 - $-$4.20)\tablenotemark{b}  & 5.8 (5.50 - 6.15)  \nl
WR 78  & WN7h    & 2.00 (1.58 - 2.00)  & 1.55 & 1375 (1365 - 1385) & $-$4.25 ($-$4.42 - $-$4.10)\tablenotemark{c}  & 5.9 (5.63 - 6.20)  \nl
WR 79a & WN9ha   & 1.99                & 1.28 & 955                & $-$4.60 ($-$4.57 - $-$4.62)\tablenotemark{d}  & 5.78  \nl
\enddata
\tablenotetext{a}{
Spectral types are from van der Hucht (2001). Data for WR 16 and WR 78  show the values adopted
in this study followed in parentheses by the range of values quoted in the literature
(Crowther et al. 1995; Nugis \& Lamers 2000; van der Hucht 2001; Hamann et al. 2006).
Data for WR 79a are from Crowther \& Bohannan (1997).}
\tablenotetext{b}{Clumping-corrected values are log $\dot{\rm M}$ = $-$4.55 at d = 3.93 kpc from radio data (Nugis \& Lamers 2000)
and log $\dot{\rm M}$ = $-$4.3  from empirical relations (Hamann et al. 2006).}
\tablenotetext{c}{Clumping-corrected values are log $\dot{\rm M}$ = $-$4.42 at d = 1.58 kpc from radio data (Nugis \& Lamers 2000)
and log $\dot{\rm M}$ = $-$4.2  from empirical relations (Hamann et al. 2006).}
\tablenotetext{d}{Radio-derived and spectroscopic values are in good agreement, and clumping effects
are thought to be small in WR 79a (Crowther \& Bohannan 1997).}
\end{deluxetable}
\clearpage
%
\begin{deluxetable}{lllllllll}
\tabletypesize{\scriptsize}
\tablewidth{0pt} 
\tablecaption{WNL Star X-ray Source Properties\tablenotemark{a}}
\tablehead{
 	   \colhead{Name}                         &
           \colhead{R.A.}                         &
           \colhead{Decl.}                        &
           \colhead{Counts\tablenotemark{b}}      &
           \colhead{Rate}                         & 
           \colhead{E$_{50}$}                     &
           \colhead{P$_{\rm cons}$}               &
           \colhead{log L$_{x}$\tablenotemark{c}} &
           \colhead{Identification(offset)}       \\
           \colhead{}                             &        
           \colhead{(J2000)}                      &
           \colhead{(J2000)}                      &
           \colhead{(cts)}                        &               
           \colhead{(cts/ks)}                     & 
           \colhead{(keV)}                        & 
           \colhead{}                             &
           \colhead{(ergs/s)}                     & 
           \colhead{(arcsec)} 
                                  }
\startdata

WR 16  & 09 54 52.94 & $-$57 43 39.2 & 31$\pm$6   &  1.03$\pm$0.20 & ...\tablenotemark{d}   & ...\tablenotemark{d} & 31.14 & GSC J095452.90$-$574338.27 (0.96)\\
WR 78  & 16 52 19.18 & $-$41 51 15.8 & 800$\pm$28 &  30.5$\pm$1.10 & 2.47                   &    0.12 (0.97)       & 32.84 & GSC J165219.25$-$415116.25 (0.93) \\          
WR 79a\tablenotemark{e} & 16 54 58.52 & $-$41 09 02.9 & 805$\pm$29 & 23.0$\pm$0.83  & 1.35  &    0.63 (0.33)       & 32.14 & GSC J165458.50$-$410903.08 (0.30) \\
\enddata
\tablenotetext{a}{
Notes:~ {\em XMM-Newton} data  are from EPIC pn.  Parameters were determined using events in
the 0.3 - 8 keV range.  Tabulated quantities are: 
source name, J2000.0 X-ray position (R.A., Decl.), pn net counts and 
net counts error (background-subtracted), 
pn count rate (Rate) obtained by dividing net counts by the usable exposure 
times given below, median photon  energy  E$_{50}$, probability of constant count-rate (P$_{\rm cons}$) 
from the KS test followed in parenthese by the value determined from $\chi^2$ analysis of 
background-subtracted binned light curves using 1000 s bins, unabsorbed X-ray luminosity L$_{x}$ (0.3 - 8 keV)  assuming
the distances in Table 2,  and HST GSC v2.3.2 optical counterpart identification.
The offset (in arcsecs) between the X-ray and optical counterpart position is given 
in parentheses. The values of E$_{50}$ and P$_{\rm const}$ were evaluated using
events within an extraction circle of radius 15$''$ (68\% encircled energy)
in order to minimize background effects. Usable pn exposure times after removing
high-background intervals were: 30.0 ks (WR 16), 26.26 ks (WR 78), and 35.03 ks (WR 79a).
}
\tablenotetext{b}{Net counts for WR 78 and WR 79a are based on background-subtracted events within an extraction circle of radius
45$''$ (90\% encircled energy). A smaller extraction region of radius 15$''$ (68\% encircled energy) was used for the faint
source WR 16 to minimize background contribution.}
\tablenotetext{c}{Unabsorbed L$_{x}$ values for WR 78 and WR 79a are based on fluxes measured in 
spectral fits after resetting the column density to zero (N$_{\rm H}$ = 0) for all model components. 
The value for WR 16 is based on a PIMMS simulation (Sec. 3.1).}
\tablenotetext{d}{Insufficient counts for a reliable measurement.}
\tablenotetext{e}{See Skinner et al (2010) for additional information on WR 79a.}
\end{deluxetable}
\clearpage
\begin{deluxetable}{ll}
\tabletypesize{\scriptsize}
\tablewidth{0pc}
\tablecaption{X-ray Spectral Fits of WR 78   
   \label{tbl-1}}
\tablehead{
\colhead{Parameter}      &
\colhead{Value }        
}
\startdata
Model                               & N$_{\rm H,1}$*kT$_{1}$ $+$ N$_{\rm H,2}$*kT$_{2}$         \nl
Abundances\tablenotemark{a}         & WN                                     \nl
N$_{\rm H,1}$ (10$^{22}$ cm$^{-2}$) & 1.38 [1.20 - 1.60]  \nl
kT$_{1}$ (keV)                      & 0.64 [0.48 - 0.88]  \nl
norm$_{1}$ (10$^{-6}$)              & 3.59 [2.40 - 6.30]  \nl 
N$_{\rm H,2}$ (10$^{22}$ cm$^{-2}$) & 6.55 [4.10 - 10.7]  \nl
kT$_{2}$ (keV)                      & 2.27 [1.63 - 2.90]   \nl
norm$_{2}$  (10$^{-6}$)             & 14.5 [8.30 - 34.4]   \nl
$\chi^2$/dof                        & 49.2/51              \nl
$\chi^2_{\nu}$                      & 0.96                 \nl
F$_{\rm X}$ (10$^{-13}$ ergs cm$^{-2}$ s$^{-1}$)         & 2.15 (14.5)     \nl
F$_{\rm X,1}$ (10$^{-13}$ ergs cm$^{-2}$ s$^{-1}$)       & 0.33 (4.27)     \nl
log L$_{\rm X}$ (ergs s$^{-1}$)     & 32.84                 \nl
log [L$_{\rm X}$/L$_{star}$]         & $-$6.64               \nl
log [L$_{\rm X}$/L$_{wind}$]        & $-$4.68              \nl
\enddata
\tablecomments{
Based on  simultaneous XSPEC (vers. 12.4.0) fit of the background-subtracted  
pn, MOS1, MOS2 spectra binned 
to a minimum of 20 counts per bin using the optically thin thermal plasma
$vapec$ model. All models included the XSPEC $wabs$ photoelectric absorption component,
based on Morrison \& McCammon (1983) cross-sections and Anders \& Ebihara (1982)
relative abundances. The tabulated parameters
are total equivalent neutral H absorption column density (N$_{\rm H}$) including
both ISM  and local stellar (e.g. wind) contributions, the product of 
Boltzmann's constant time plasma temperature  (kT),
and XSPEC component normalization (norm). 
Square brackets enclose 90\% confidence intervals.
The total X-ray flux (F$_{\rm X}$) and flux of the cool component
(F$_{\rm X,1}$) are the absorbed values in the 0.3 - 8 keV range, followed in 
parentheses by  unabsorbed values. 
The unabsorbed luminosity L$_{\rm X}$ (0.3 - 8 keV)  assumes
d = 1.99 kpc (Table 2).   L$_{star}$ and 
L$_{wind}$ = (1/2)$\dot{\rm M}$v$_{\infty}^2$ are from Table 2.
}
\tablenotetext{a}{Abundances were held fixed at the generic WN 
values given in Table 1 of Van der Hucht et al. (1986).
The generic WN abundances reflect H depletion and N enrichment and
are by number:
He/H = 14.9, C/H = 1.90E-03, N/H = 9.36E-02,
O/H = 4.35E-03, Ne/H = 9.78E-03, Mg/H = 3.26E-03, 
Si/H = 3.22E-03, P/H = 1.57E-05, S/H = 7.60E-04,
Fe/H = 1.90E-03. All other elements were held fixed
at solar abundances (Anders \& Grevesse 1989). }

\end{deluxetable}
\clearpage

\begin{figure}
\figurenum{1}
\includegraphics*[width=7.0cm,angle=-90]{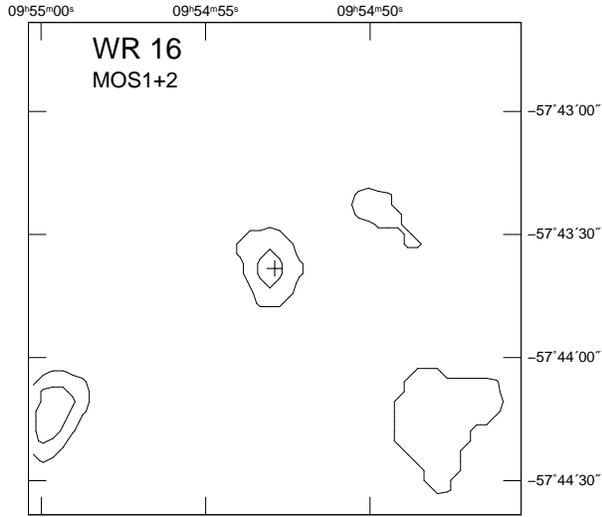}
\caption{Summed EPIC MOS1$+$2 contour image  of WR 16
(0.3 - 8 keV; 33 ks of usable exposure per MOS; Gaussian-smoothed). The inner contour 
(2$\sigma$) encloses a weak excess at the HST GSC stellar position (cross).}
\end{figure}

\vspace*{0.5in}

\begin{figure}
\figurenum{2}
\includegraphics*[width=7.0cm,angle=-90]{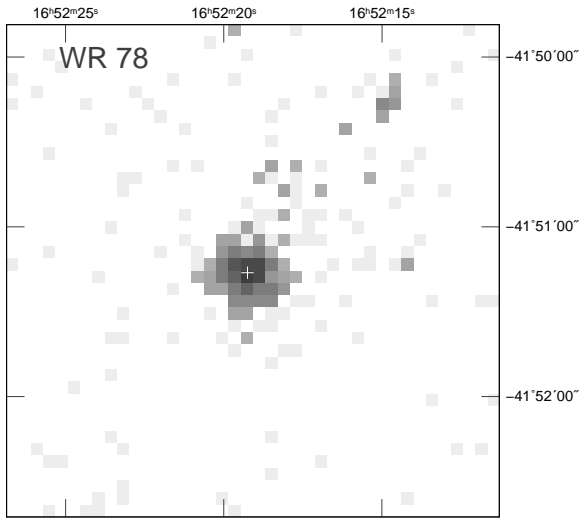}
\caption{EPIC pn image  of WR 78
(0.3 - 8 keV; 26.3 ks of usable exposure; 4.$''$3 pixels) showing a clear
detection  at the HST GSC stellar position (cross).}
\end{figure}


\clearpage

\begin{figure}
\figurenum{3}
\includegraphics*[width=7.0cm,angle=-90]{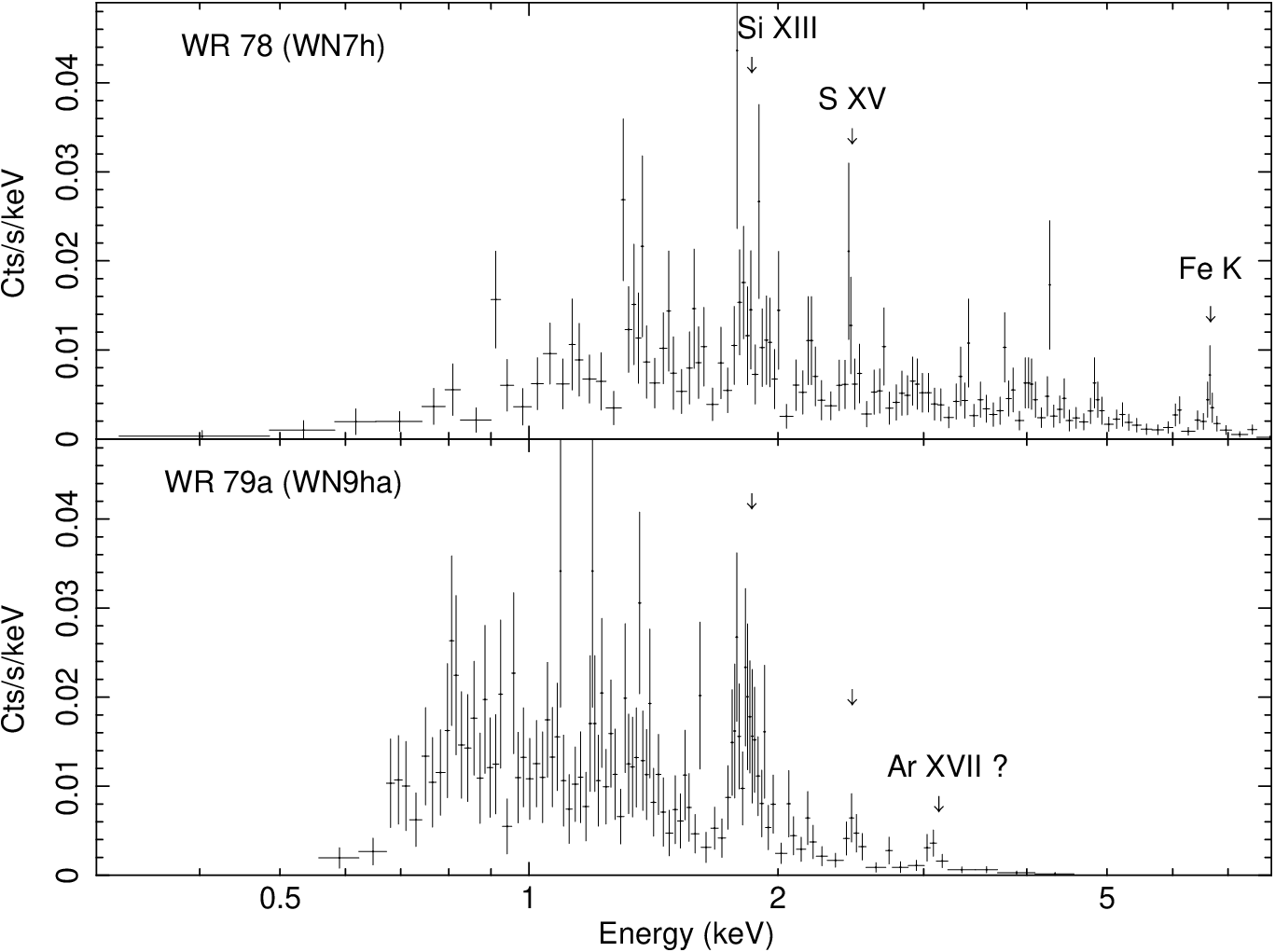} \\
\includegraphics*[width=7.0cm,angle=-90]{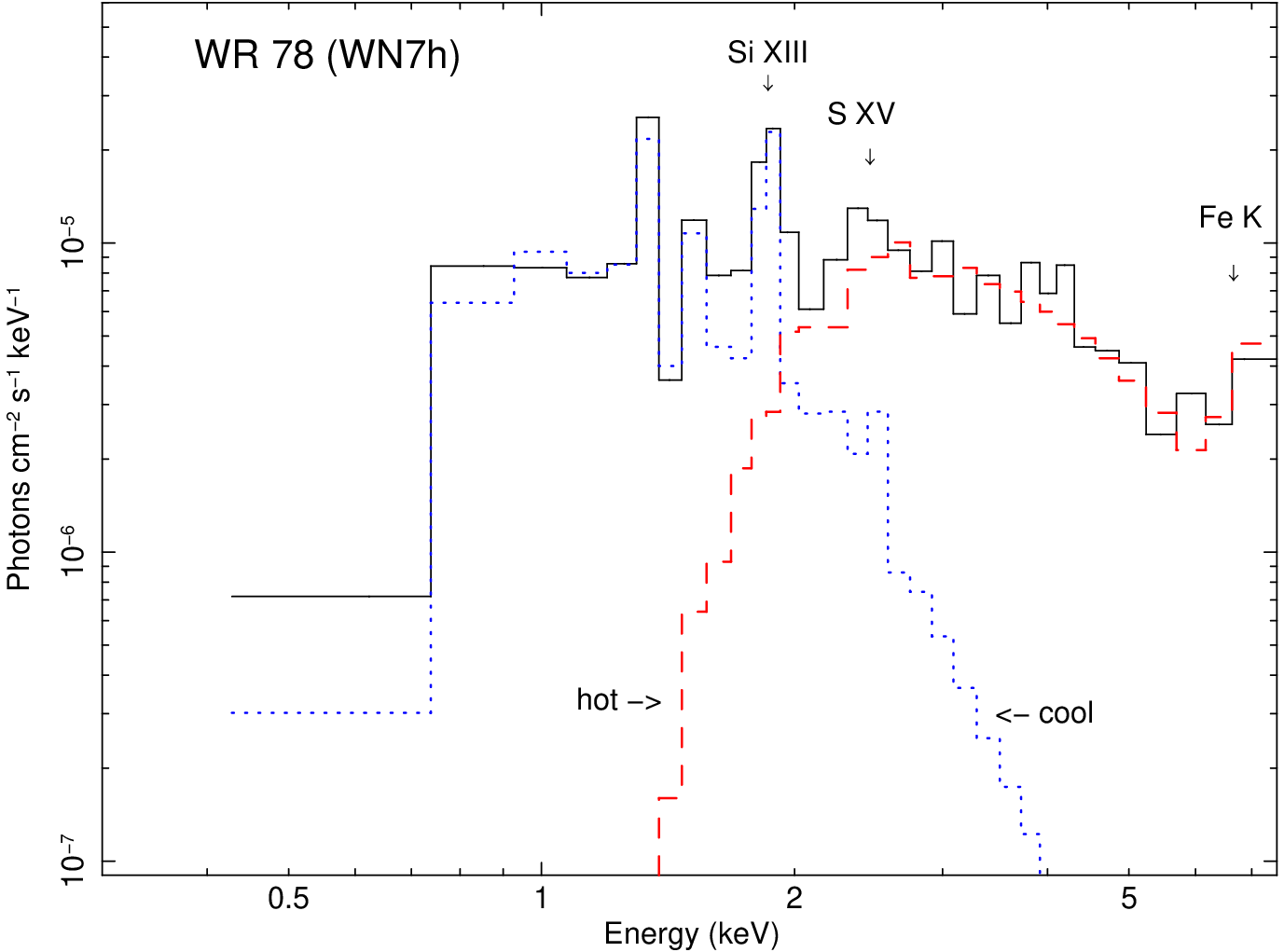}
\caption{Top and Middle:~EPIC pn background-subtracted spectra   of WR 78 and
WR 79a, binned to a minimum of 5 counts per bin. Spectral fit
parameters for WR 78 are given in Table 4. Best-fit parameters
for WR 79a  are N$_{\rm H}$ = 7.5 [6.4 - 8.3]
$\times$ 10$^{21}$ cm$^{-2}$,
kT$_{1}$ = 0.55 [0.46 - 0.61] keV,
and kT$_{2}$ = 2.66 [1.66 - 5.60] keV where brackets
enclose 90\% confidence ranges (S10).~
Bottom: Unfolded spectrum of WR 78 binned to a minimum of
20 counts per bin showing the separate contributions
of the cool (dotted line) and hot (dashed line) components (Table 4). 
The hot component accounts for most of the emission above 2 keV
including the high-temperature Fe K line.
Error bars omitted for clarity; log-log scale.
}
\end{figure}

\clearpage

\begin{figure}
\figurenum{4}
\includegraphics*[width=7.0cm,angle=-90]{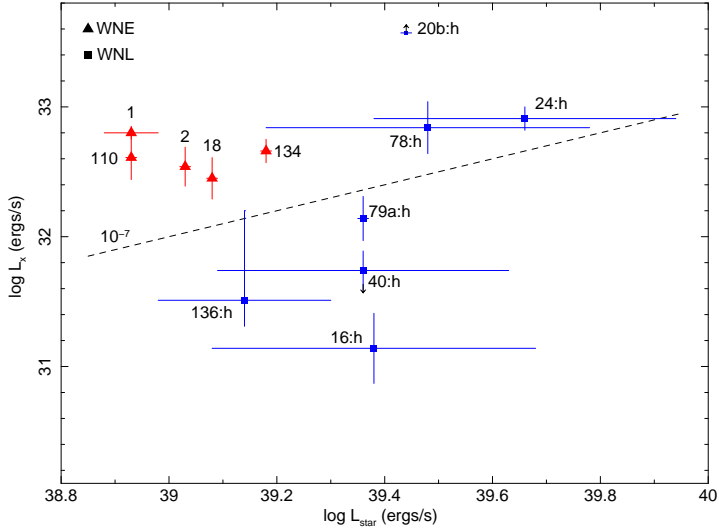}
\caption{Unabsorbed X-ray luminosity (0.3 - 8 keV) versus 
stellar  luminosity for WN stars. 
Assumed distances are given in Table 2 and S10.
WR star numbers are from van der Hucht 2001 and :h
following the number denotes the presence of hydrogen in the spectrum.
WNE and WNL classifications follow Hamann et al. 2006.
Unabsorbed L$_{x}$ was obtained by setting the column density to 
zero (N$_{\rm H}$ = 0) in best-fit spectral models and measuring 
the unabsorbed flux. Exceptions are the faint source WR 16 (PIMMS
simulations; Sec. 3.1) and the undetected star WR 40.
Unabsorbed L$_{x}$ values are from this work (WR 16, WR 78),
Skinner et al. 2002 (WR 110),  and Skinner et al. 2010 (and 
references therein). The L$_{x}$ value for WR 1  is based on 
analysis of {\em XMM} archive data (obsid 0552220101) and assumes 
d = 1.82 kpc (van der Hucht 2001).  The L$_{x}$ upper limit for WR 40 is from 
Gosset et al. 2005,  adjusted upward by 0.14 dex to allow for distance 
uncertainties. L$_{star}$ uncertainties 
reflect the range of  values published in the literature (Crowther et al. 1995; 
Crowther \& Bohannan 1997; Nugis \& Lamers 2000; Hamann et al. 2006).
The dashed line shows the canonical L$_{x}$ = 
10$^{-7}$L$_{bol}$ relation for O and early B-type stars.
The spectral types from van der Hucht (2001) are:
WR 1 (WN4), WR 2 (WN2), WR 16 (WN8h), WR 18 (WN4), WR 20b (WN6:h),
WR 24 (WN6ha), WR 40 (WN8h), WR 78 (WN7h), WR 79a (WN9ha),
WR 110 (WN5), WR 134 (WN6), WR 136 (WN6h). 
}
\end{figure}

\clearpage

\begin{figure}
\figurenum{5}
\includegraphics*[width=7.0cm,angle=-90]{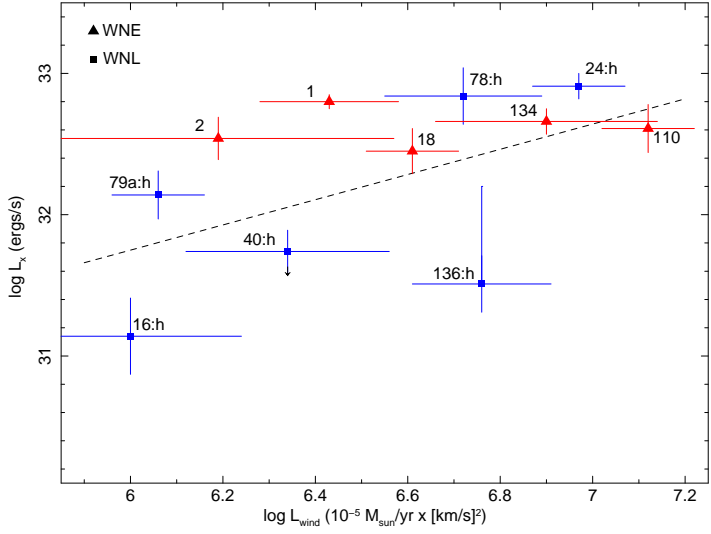}
\caption{Unabsorbed X-ray luminosity (0.3 - 8 keV)  versus wind luminosity 
L$_{wind}$ =   $\frac{1}{2}$$\dot{\rm M}$v$_{\infty}^2$
for  WN stars. Mass-loss data are based on published values (see references below) and
uncertainties reflect the range of published values. WR 20b is excluded for lack of
reliable mass-loss data. The dashed line shows a linear regression fit 
log L$_{x}$ = 0.89($\pm$ 0.47)log L$_{wind}$ $+$ 26.41($\pm$0.55)  ergs s$^{-1}$.
Symbols and spectral types are the same as given in Figure 4.
Mass-loss parameter references (in parentheses):~
WR 1:~(4,7,10)~
WR 2:~(1,2,4,8)~
WR 16:~(3,4,10);
WR 18:~(4);
WR 24:~(3,4,10);
WR 40:~(3,4,9,10)
WR 78:~(3,4,10);
WR 79a:~(5);
WR 110:~(4);
WR 134:~(1,4,10);
WR 136:~(1,2,4,6,7,10).~{\em References}:
(1) Abbott et al. 1986;
(2) Howarth \& Schmutz 1992;
(3) Crowther et al. 1995;
(4) Hamann et al. 2006;
(5) Crowther \& Bohannan 1997;
(6) Ignace et al. 2001;
(7) Ignace et al. 2003b;
(8) van der Hucht 2001;
(9) Gosset et al. 2005;
(10) Nugis \& Lamers 2000.
}
\end{figure}

\end{document}